\documentclass[pdflatex,sn-mathphys-num]{sn-jnl}

\usepackage{graphicx}%
\usepackage{multirow}%
\usepackage{amsmath,amssymb,amsfonts}%
\usepackage{amsthm}%
\usepackage{mathrsfs}%
\usepackage[title]{appendix}%
\usepackage{xcolor}%
\usepackage{textcomp}%
\usepackage{manyfoot}%
\usepackage{booktabs}%
\usepackage{algorithm}%
\usepackage{algorithmicx}%
\usepackage{algpseudocode}%
\usepackage{listings}%
\usepackage{dsfont}
\usepackage[version=4]{mhchem}

\theoremstyle{thmstyleone}%
\theoremstyle{thmstyletwo}%

\theoremstyle{thmstylethree}%

\begin{document}

\title[Article Title]{Symmetry-aware generative design of flat-band materials beyond known crystal-net prototypes}


\author[1]{\fnm{Yihao} \sur{Wei}}

\author[1]{\fnm{Ivan} \sur{Savochkin}}
\author[1]{\fnm{Artem} \sur{Mishchenko}}
\author*[1]{\fnm{Xiangwen} \sur{Wang}}\email{xiangwen.wang@manchester.ac.uk}

\author*[1]{\fnm{Qian} \sur{Yang}}\email{qian.yang@manchester.ac.uk}

\affil[1]{\orgdiv{Department of Physics and Astronomy}, \orgname{The University of Manchester}, \orgaddress{\street{Oxford Road}, \city{Manchester}, \postcode{M13 9PL}, \country{UK}}}


\abstract{Flat electronic bands underlie a range of strongly correlated and topological phenomena, whose design in real materials has so far relied on a small catalogue of named geometric motifs such as kagome, Lieb, and pyrochlore nets.  This discrete catalogue is by no means to exhaust the geometries that support flat bands in real compounds, as band flatness is a property of network connectivity. Here we combine a continuous geometric representation of crystal sublattices, with a symmetry-constrained generative model, to access a broader design space for materials hosting flat bands. The key step is to choose sublattice motifs that are outside the known geometric clusters, ensuring the novelty of the generated structures. We then introduce SkeleGen, which pins these unconventional skeletons to symmetry-compatible Wyckoff positions while denoising the surrounding chemistry, resulting in 9,352 crystal candidates that survive stability and flatnessscreening. Band flatness is confirmed using high throughput full DFT calculations, which agree well also with the tight-binding spectra of the isolated skeletons, supporting a geometric origin of the band flatness. We demonstrate "out-of-distribution" motifs as a new design principle to dramatically expand geometric repertoire for materials discovery, potentially beyond flat bands. }



\maketitle

\section*{Introduction}\label{intro}

Flat electronic bands underlie a range of strongly correlated phenomena, including unconventional superconductivity, fractional Chern insulators, and ferromagnetism \cite{heikkila2011flat, tang2011high, sun2011nearly, balents2020superconductivity, miyahara2007bcs, lin2018flatbands}. Their realization in twisted bilayer graphene, kagome metals, and Lieb-lattice oxides has established flat-band engineering as a central strategy in the search for correlated quantum materials~\cite{checkelsky2024flat, bistritzer2011moire, cao2018unconventional, park2023topological, lin2018flatbands, zhang20252d, mukherjee2015observation, slot2017experimental}. In these systems, quenching electronic kinetic energy amplifies the role of electron-electron interactions, making flat bands a powerful platform for designing quantum phases that are difficult to access in dispersive bands.

A  major class of flat bands arises as a collective consequence of network connectivity, in which destructive interference produces compact localized states and, consequently, dispersionless bands. For example, the flat bands in well-known geometry motifs, such as kagome, Lieb, dice, line-graph bipartite networks, and three-dimensional analogues such as pyrochlore, are all attributed to geometric connectivity ~\cite{jacqmin2014direct, regnault2022catalogue, cualuguaru2022general, mukherjee2015observation, wang2011nearly, trescher2012flat, maimaiti2017compact, flach2014detangling}. The search for flat-band materials has therefore predominantly targeted this handful of known motifs \cite{regnault2022catalogue, liu2021screening, kollar2020line}, with workflows often beginning by specifying a named motif type. However, the standard taxonomy of crystal nets~\cite{delgado2003identification,yaghi2003reticular} partitions geometries into discrete prototypes, offering no natural means of identifying structures that fall between them. Consequently, this approach leaves large regions of geometric space with unconventional motifs unexplored, as they are difficult to access in practice. Materials whose flat bands arise from connectivity patterns outside the named families are therefore effectively invisible to current design strategies, even when they are already present in existing structural databases.

Recent progress in generative deep learning offers a route around this bottleneck beyond enumerating known prototypes, by proposing candidate structures directly. Variational autoencoders, generative adversarial networks, and autoregressive models have been applied to crystal 
generation~\cite{court20203,ren2022invertible,nouira2018crystalgan,kim2020generative}, and diffusion-based frameworks have since become the dominant paradigm, with CDVAE~\cite{xie2021crystal}, DiffCSP~\cite{jiao2023crystal}, MatterGen~\cite{zeni2025generative}, and related models achieving substantial gains in structural validity and property control. However, the success of these models is usually evaluated in terms of structural validity, stability, novelty, or target-property satisfaction. These objectives do not specify how a particular connectivity pattern should be retained or how the electronic feature of interest should remain tied to that pattern after generation. Recent symmetry- and structure-aware models have started addressing this problem. DiffCSP++~\cite{jiao2024space} enforces space-group symmetry through constrained diffusion in an invariant lattice representation and Wyckoff-position-constrained coordinate updates. However, its constraints act on the global crystallographic degrees of freedom of the generated structure rather than externally specified local connectivity pattern. SCIGEN~\cite{okabe2025structural} introduces a masking strategy that pins a subset of atoms to a predefined geometric template throughout the diffusion trajectory, but the pinned template is treated as a structural constraint rather than as part of a space-group-consistent Wyckoff configuration. Existing methods therefore do not yet support transferring an uncatalogued flat-band connectivity pattern into a new crystal while preserving both its symmetry and electronic origin.

\begin{figure}
    \centering
    \includegraphics[width=\linewidth]{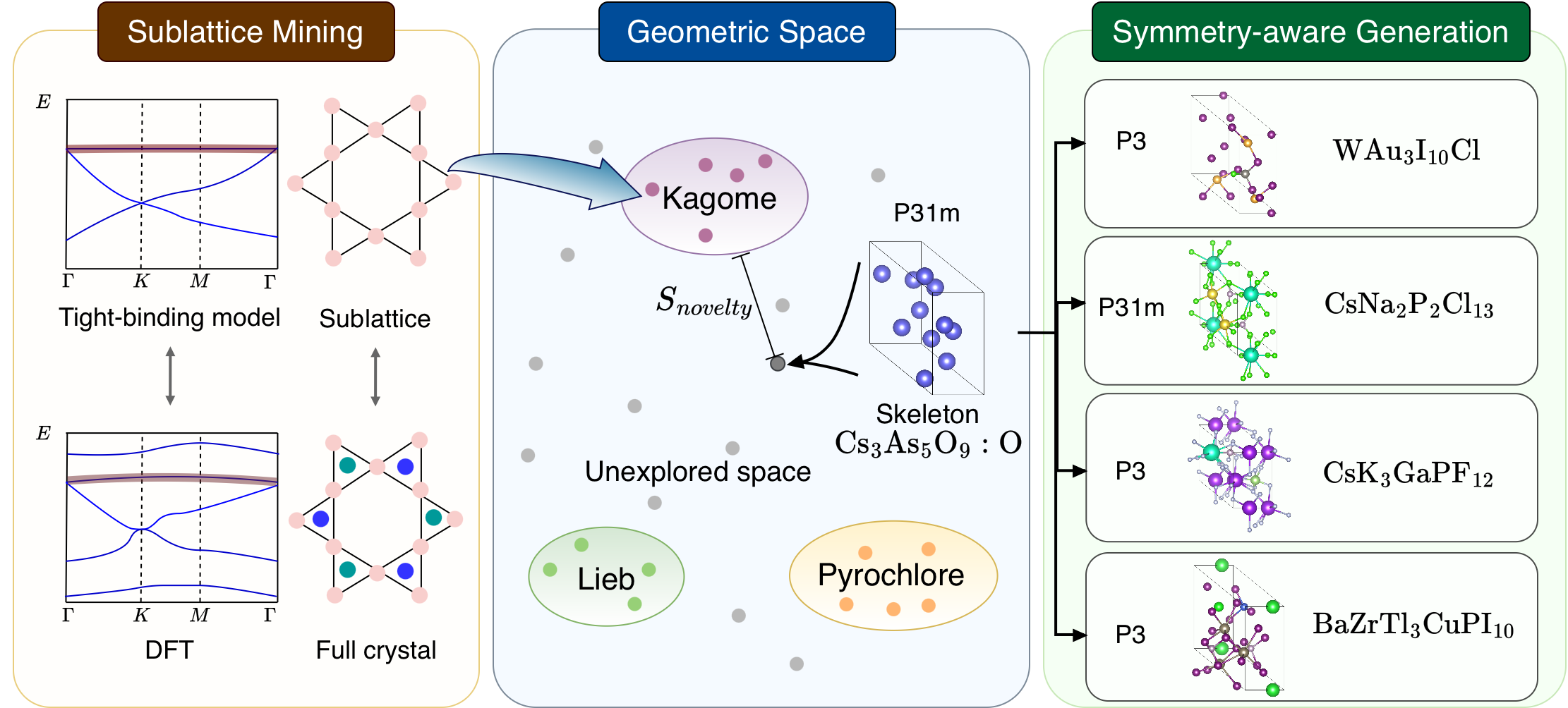}
    \caption{\textbf{Unconventional flat-band geometries and their realization.} Left: stripping a flat-band material to its bare sublattice, a nearest-neighbour tight-binding model reproduces the dispersionless band of the full DFT calculation, indicating that band flatness originates from network connectivity. Centre: validated flat-band sublattices embedded in a continuous geometric space form known clusters (kagome, Lieb, pyrochlore) separated by a sparsely populated unexplored region, where gray points represent unclustered sublattices; outliers are ranked by a novelty score $S_{\text{novelty}}$, the latent-space distance to the nearest known family. Right: an unconventional skeleton (e.g. the oxygen sublattice of \ce{Cs3As5O9}) fixed as a rigid scaffold yields chemically diverse flat-band candidates that reduce its space-group symmetry.}
    \label{fig:highlights}
\end{figure}

Here we present a skeleton-guided generative framework (SkeleGen) that turns uncatalogued flat-band geometry into a transferable design variable for crystal generation (Figure \ref{fig:highlights}). The framework identifies non-trivial flat-band skeletons in known compounds via tight-binding validation, maps them into a continuous geometric space, and selects high-novelty skeletons beyond named-net families. Our symmetry-aware skeleton generation method then converts them into symmetry-compatible Wyckoff constraints for generative crystal design. Our approach generated over one million candidates, of which 9,352 survive stability and flatness screening. High-throughput density functional theory (DFT) evaluation confirms that 73\% of scored candidates host flat bands at the targeted flatness level, with representative cases such as \ce{KNaTlCoSbF12}, \ce{La(FeO4)2}, and \ce{TaAsSF15} retaining flat features directly traceable to the imposed skeletons. Together, these results establish a route to expand flat-band materials discovery beyond known prototypes.

\section*{Results}\label{result}

\begin{figure}
    \centering
    \includegraphics[width=\linewidth]{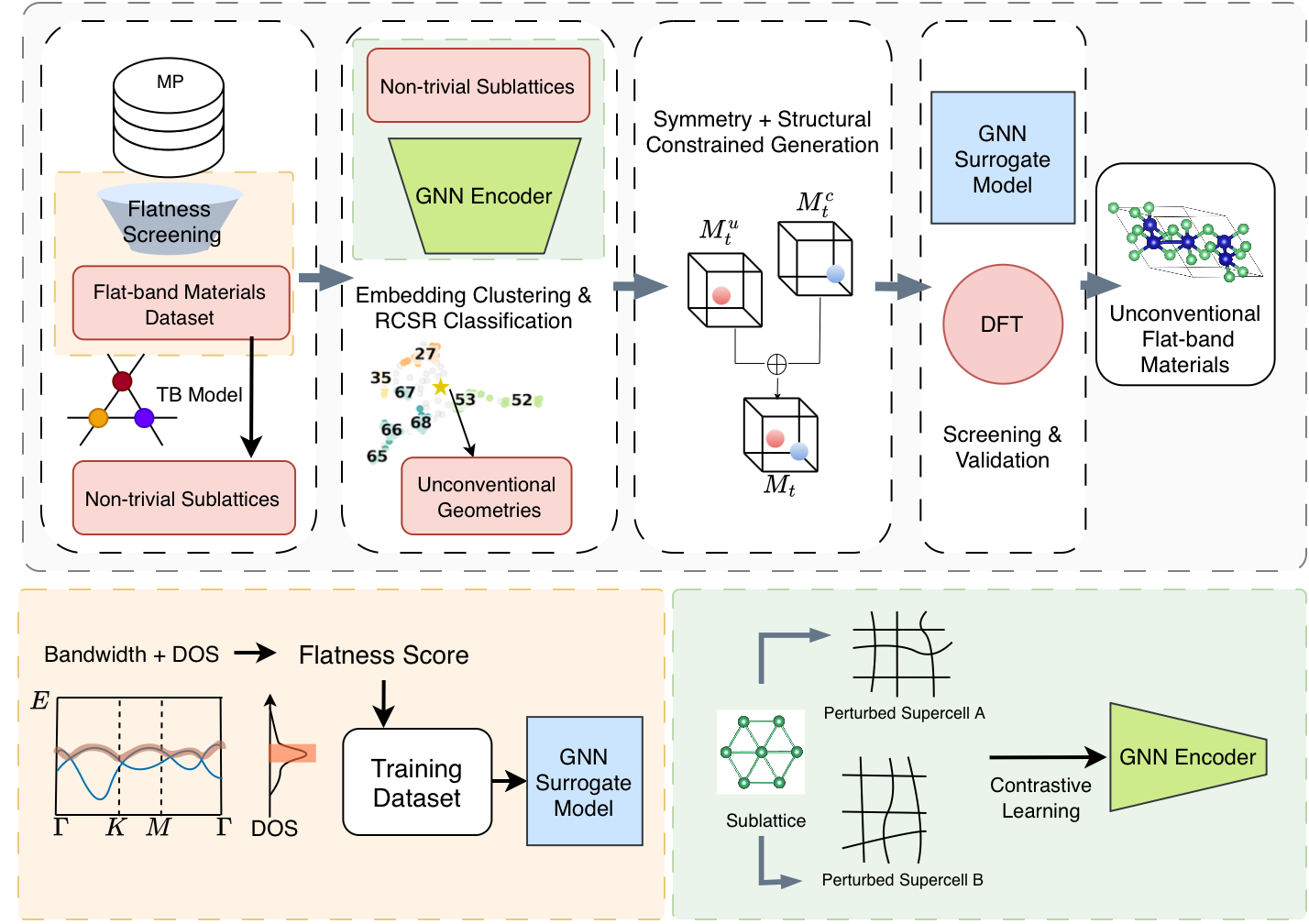}
    \caption{\textbf{Workflow of unconventional flat-band material generation.} The overall pipeline is shown in the top panel. Starting from the Materials Project (MP) database, flat-band screening and tight-binding analysis identify flat-band materials and extract non-trivial sublattices, whose GNN-encoder embeddings are clustered and classified by RCSR prototypes to discover unconventional geometries. These motifs are expanded through symmetry- and structural-constrained generation, in which the constrained skeleton $M_t^c$ and the unconstrained component $M_t^u$ are recombined at each diffusion step into the full structure $M_t$. Candidates are then screened by the GNN surrogate model and validated with DFT, yielding flat-band materials. The bottom-left panel shows surrogate-model training: a flatness score\cite{wang2025structureinformedlearningflatband} from the electronic bandwidth and density of states (DOS) labels a dataset used to fit the GNN surrogate model. The bottom-right panel shows encoder training: each crystal is decomposed into sublattices, and perturbed supercells from the same sublattice form a contrastive-learning scheme that trains the GNN encoder to produce sublattice embeddings.}
    \label{fig:pipeline}
\end{figure}

We establish unconventional crystal geometry as a generative handle for flat-band materials discovery. The central object is the flat-band skeleton -- a sublattice extracted from a real crystal whose connectivity alone produces a non-trivial flat band. A latent geometry map compares skeletons by geometric distance, exposing novel geometries outside established motif families. In the SkeleGen method, the selected skeletons are then recast from Cartesian sublattice positions into Wyckoff constraints, so that new chemical environments are generated around an intact, crystallographically embedded core. The workflow is summarized in Figure \ref{fig:pipeline}.

\subsection*{Connectivity-driven flat-band skeletons abound in real crystals}\label{struct2flat}

We constructed a dictionary of connectivity-driven flat-band sublattices from existing inorganic compounds, to better navigate the flat-band geometry motifs. Starting from the structures with electronic band information in Materials Project database~\cite{horton2025accelerated, jain2013commentary}, we identified flat-band candidates using a physics-motivated flatness score considering both bandwidth and density of states \cite{wang2025structureinformedlearningflatband}. Among 29,465 screened materials, 4,155 of them reached a flatness score above 0.95 and were retained for further sublattice-level analysis.

The presence of a narrow electronic band in a crystal does not guarantee a geometric origin; it may arise from trivial atomic insulation. To isolate flat bands due to geometric drivers, i.e., the specific connectivity motifs that induce destructive interference, we implemented a sublattice abstraction pipeline introduced by \cite{neves2024crystal}, see details in the Methods. Briefly, for each high-score material, elemental sublattices were extracted and converted into nearest-neighbour crystal graphs. We then assigned a single $s$ orbital to each site and retained only those sublattices whose nearest-neighbour Hamiltonian supports a dispersion-less band after removing isolated finite clusters. This filtering procedure identifies a dictionary of 1,852 non-trivial flat-band sublattices whose electronic flatness can be reproduced in a simplified connectivity model. Because these skeletons are extracted from DFT-characterized crystals, they retain realistic bond lengths, coordination environments and periodic embeddings. They therefore serve as a physically grounded starting point for exploring flat-band geometries beyond the known motif catalogue.

\subsection*{A continuous geometric representation reveals novel skeletons}\label{gnnencoder}

The 1,852 validated skeletons raise a question that named-net classification is poorly equipped to answer: how many of them belong to the established flat-band families, and how many fall outside? The standard taxonomy partitions geometry into discrete, catalogued prototypes and offers no measure of the space between them, so a skeleton that does not match a catalogued net is left unclassified rather than recognised as novel. To examine the question on its own terms, we replace categorical classification with a latent representation in which proximity reflects geometric kinship. To this end, a graph neural network encoder is trained by contrastive self-supervision\cite{oord2019representationlearningcontrastivepredictive} on these sublattices, using only Cartesian coordinates as node features so that the embedding reflects geometry alone, independent of chemistry (Methods). Applied to the 1,852 flat-band skeletons, this representation allows named, recurrent, and geometry outliers to be compared on the same footing.

\begin{figure}
    \centering
    \includegraphics[width=\linewidth]{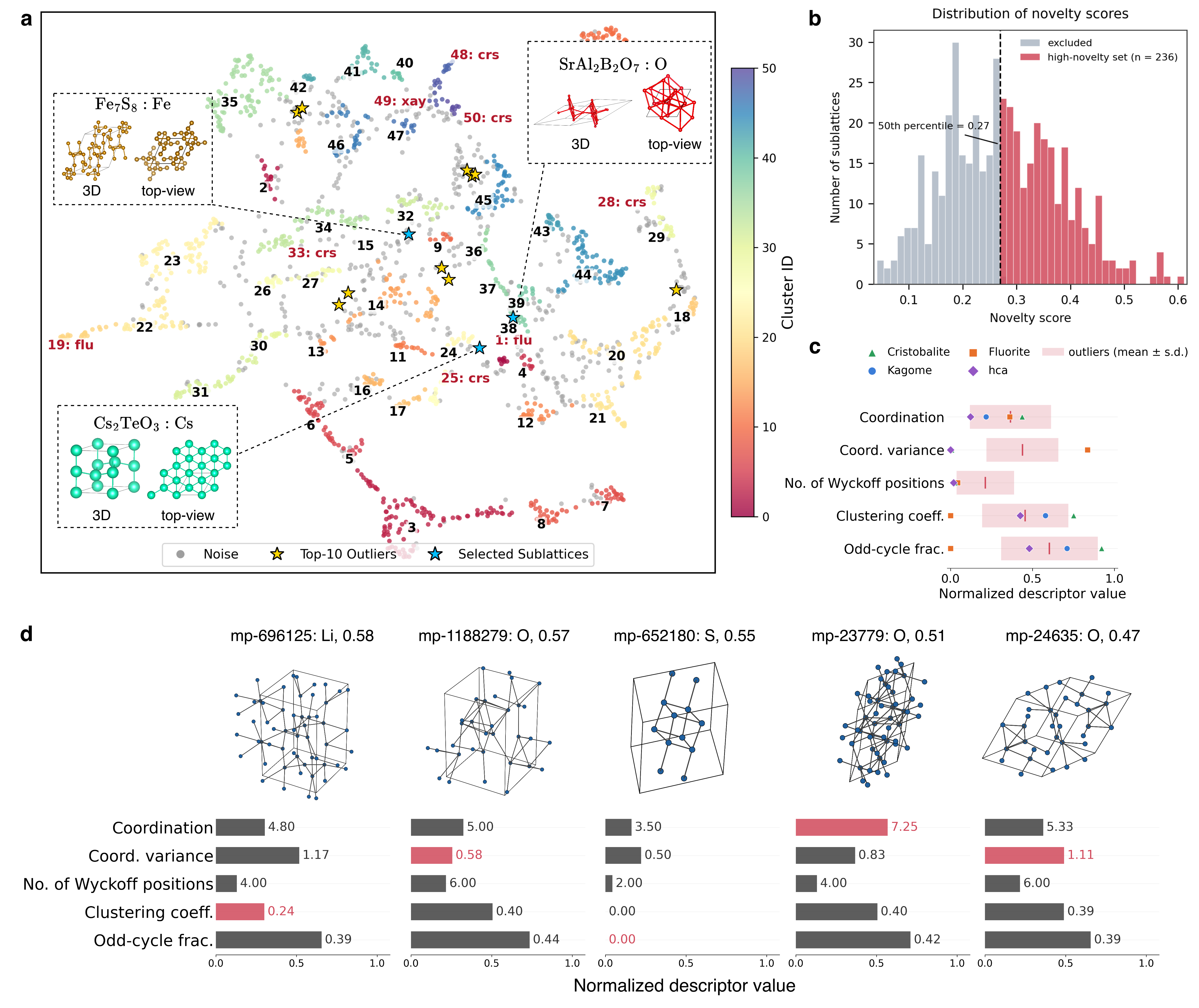}
    \caption{\textbf{Geometric latent space analysis of flat-band sublattices and identification of unconventional motifs.}
    \textbf{a}, UMAP projection of GNN-encoded latent embeddings for 1,852 non-trivial flat-band sublattices, colored by HDBSCAN cluster ID. Clusters that coincide with catalogued crystal net prototypes are labeled (e.g., flu: fluorite, crs: cristobalite and xay: deltahedron); the remaining clusters recur across the database but are not dominated by certain catalogue. Top-10 novelty outliers (yellow stars) and representative outliers (blue stars) are highlighted; grey points denote unclustered or catalogued points. Insets show three-dimensional and top-view visualizations of representative outlier sublattices extracted from their parent compounds (\ce{Fe7S8}:Fe, \ce{Cs2TeO3}:Cs, \ce{SrAl2B2O7}:O).
    \textbf{b}, Distribution of novelty scores. Sublattices with near-zero scores ($<0.01$), corresponding to points residing within or immediately adjacent to known clusters, are excluded; the dashed line marks the 50th percentile of the remaining distribution (0.27). The $n=236$ sublattices above this threshold (red) define the high-novelty set used for constrained generation.
    \textbf{c}, Five normalized graph-theoretic descriptors -- mean coordination number, coordination variance, number of Wyckoff positions, clustering coefficient and odd-cycle fraction -- for representative named families (kagome, fluorite, cristobalite, hca; markers) and for the high-novelty outlier population (mean $\pm$ 1~s.d., shaded band). 
    \textbf{d}, Six selected high-novelty sublattices (Materials Project ID, skeleton element and novelty score above each panel; three-dimensional net rendering) with their five descriptor values below. Bar length gives the normalized descriptor value and the raw value is annotated; for each sublattice the descriptor on which it departs most strongly from the average is highlighted in red.}
    \label{fig:latent_space}
\end{figure}

Identification of known prototypes is performed through barycentric placement algorithm with Systre\cite{delgado2003identification}, which assigns a catalogued RCSR name\cite{yaghi2003reticular} to 334 of the 1,852 skeletons. This small fraction reflects both the finite size of the named catalogue and the limited ability of barycentric placement to classify low-symmetry, distorted, or three-dimensionally embedded sublattices that are common among real crystals. A density-based pass with HDBSCAN\cite{mcinnes2017hdbscan}, which does not require a catalogued match, groups 1,295 of the skeletons into clusters of recurrent geometry, as visualized in Figure \ref{fig:latent_space}\textbf{a}. The two references diverge substantially: only a few clusters coincide with catalogued families, including kagome-like networks, Star sublattices, and fluorite-type nets, while most clusters correspond to geometries that recur across the database but carry no catalogue name, and some catalogued labels are dispersed by real-crystal distortions and three-dimensional embedding. The named-motif catalogue therefore describes only a small fraction of the flat-band geometries presented in crystals, while even the recurrent, frequently occurring geometries are largely uncatalogued. Such mismatch provides the first evidence that connectivity-driven flat bands extend well beyond the named families.

Beyond the recurrent regions lies a further population that is neither named nor clustered. To quantify how far a skeleton sits from the geometries that are already accounted for, we define a novelty score as the minimum latent-space distance to any Systre-classified or HDBSCAN-clustered point. After discarding near-zero scores (below 0.01), corresponding to skeletons that sit within or immediately adjacent to a reference region, we retain 236 skeletons above the 50th-percentile threshold (novelty score 0.27) as candidates for further generation (Figure~\ref{fig:latent_space}\textbf{b}). This score measures the separation from known regions, but it does not describe the internal structure of the outlier set. To do this, we re-express each skeleton using interpretable graph-theoretic descriptors (Figure~\ref{fig:latent_space}\textbf{c}). Compared with the narrow, characteristic profiles of the named families (symbols, colour coded), the outlier population (pink bars) spans broader ranges that entail every family across all five descriptors. We show five individual high-novelty outliers, each occupies distinct regions of the descriptor space, from networks with no closed odd cycles (clustering coefficient and odd-cycle fraction = 0) to densely coordinated, odd-cycle-rich networks (Figure~\ref{fig:latent_space}\textbf{d}, more details in SI III). These outliers therefore form a geometrically diverse set of novel flat-band skeletons beyond named-net classification.

The 236 high-novelty skeletons are therefore unified not by a shared geometry, but by a shared invisibility to naming-based design: each escapes the standard catalogue, though for different reasons. We exemplify three categories from these high novelty skeletons. The first is a three-dimensional expansion of a well-studied lower-dimensional net, where the added dimensionality introduces deviations large enough to prevent classification. For example, the Cs sublattice of \ce{Cs2TeO3} (mp-614803) is representative, forming an approximate triangular net when viewed from the out-of-plane direction while its nominally in-plane atoms do not lie strictly in the same plane. The second is a known flat-band motif embedded within a complex local environment that obscures it, as in the oxygen sublattice of \ce{SrAl2B2O7} (mp-15939), where a corner-sharing tetrahedral geometry sits amid numerous decorative atoms. The third contains connectivity patterns that cannot be referred to any known flat-band geometries, such as the Fe sublattice in \ce{Fe7S8} (mp-542794). None of these geometries could  be easily implemented into a design workflow that begins with a named prototype, but become accessible in our workflow as flatness-hosting connectivity is identified directly. Below, we demonstrate the constrained generation using these skeletons. 

\subsection*{Novel skeletons as symmetry-compatible generative constraints}\label{con_gen}

Having isolated a subset of 236 novel, non-trivial flat-band sublattices, we then use them as the skeleton to expand the design space of flat-band quantum materials. A major task is to fill such a fixed geometric core with different chemistry environment, while keeping the core intact. This implies that the core must be kept intact not only geometrically, but also crystallographically. The topological structure that makes a flat band particularly interesting, including its band touching with neighbouring dispersive bands and the associated Berry-phase character, is symmetry dependent \cite{cualuguaru2022general, setty2024symmetry, khalaf2022symmetry}. A skeleton transplanted into new chemistry could only retain these features if its symmetry is tracked consistently throughout generation. While adding unconstrained atoms will only lower or preserve the symmetry of the skeleton, the generated crystal in general belong to a subgroup of the skeleton's space group. Therefore, the skeleton is required to be placed within a crystallographically well-defined space group throughout generation.

\begin{figure}
    \centering
    \includegraphics[width=\linewidth]{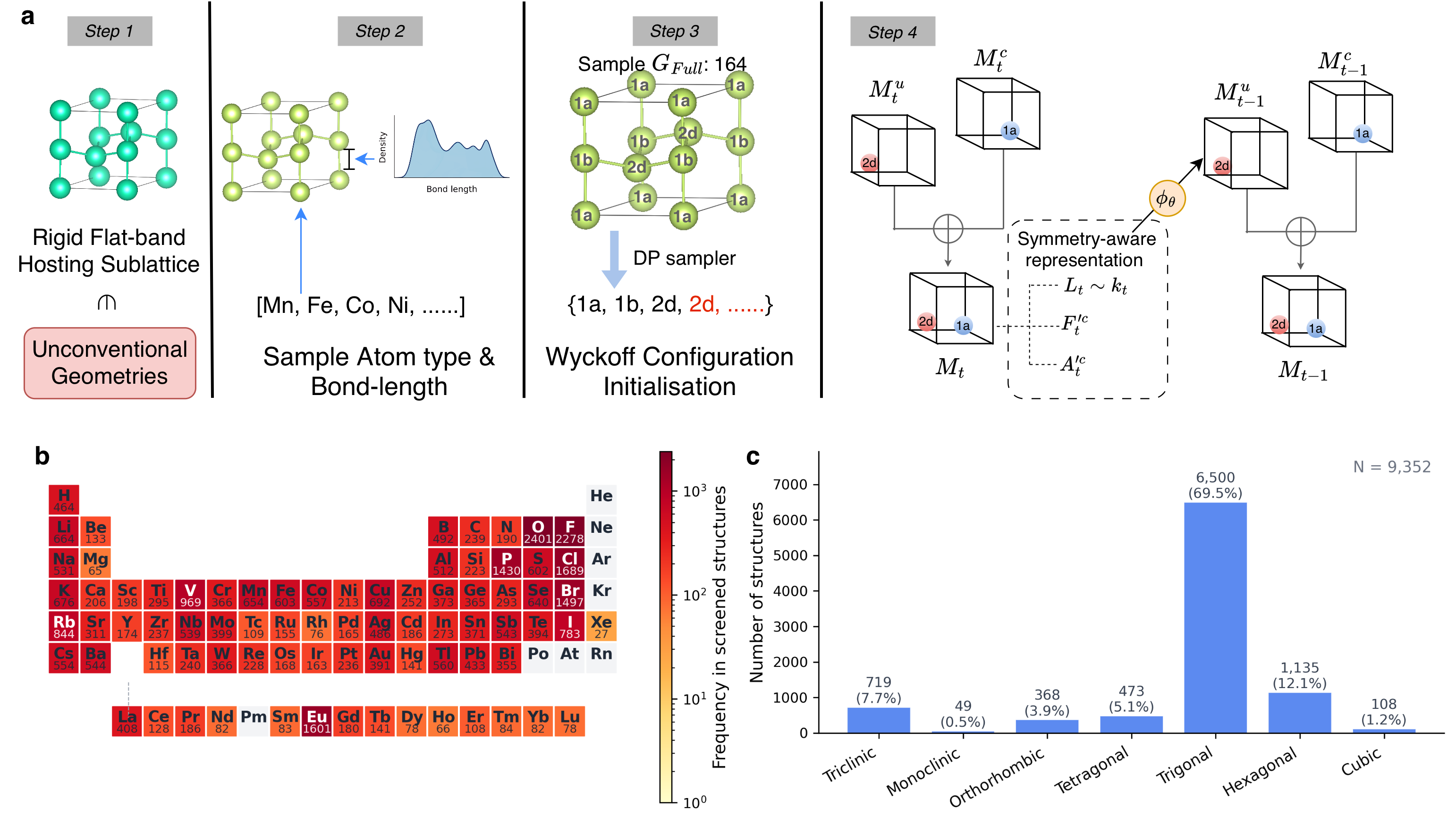}
    \caption{\textbf{SkeleGen framework and statistics of screened candidate materials.}
    \textbf{a},~Overview of the constrained generation pipeline.
    Step~1: a rigid flat-band-hosting sublattice with unconventional geometry is selected as the structural skeleton.
    Step~2: sublattice sites are populated with predetermined species and lattice lengths are sampled from bond-length distributions.
    Step~3: a compatible space group $G_{\text{full}}$ is sampled, constrained atoms are matched to Wyckoff positions, and the remaining positions are completed by a dynamic-programming sampler, yielding a full Wyckoff configuration.
    Step~4: DiffCSP++ performs constrained diffusion in which the skeleton atoms follow a predefined trajectory ($\mathcal{M}_t^c$) while unconstrained atoms ($\mathcal{M}_t^u$) are denoised by the model $\phi_\theta$; the two components are recombined at each timestep via the constraint mask.
    \textbf{b},~Periodic-table heatmap of elemental frequency across screened candidate structures, shown on a logarithmic color scale. Transition metals and chalcogenides dominate the chemical space.
    \textbf{c},~Distribution of screened structures ($N = 9{,}352$) across the seven crystal systems, with counts and percentages annotated.}
    \label{fig:scigenp}
\end{figure}

We develop a symmetry-aware skeleton generation method called SkeleGen, where the skeleton is recast from a set of fixed fractional coordinates into a symmetry constraint (Figure \ref{fig:scigenp}\textbf{a}; Methods; SI IV). Skeleton sites are populated by sampling broadly from transition metals, lanthanides, main-group metals, and non-metals; lattice angles are inherited from the sublattice, while lattice lengths are sampled from Materials Project bond-length distributions. Applying SkeleGen to these 236 novel skeletons produces 1,030,975 candidate crystals, demonstrating that unconventional flat-band geometries can be embedded into a large chemical and structural search space. After a four-stage stability screening comprising SMACT charge-balance, occupation-ratio, energy-above-hull, and non-diffused-structure checks, similar to that in SCIGEN\cite{okabe2025structural}, 72,544 candidates remain.

\subsection*{Generated crystals retain flat bands traceable to imposed skeleton}\label{validation}

The remaining 72,544 candidate materials are far too many for direct first-principles evaluation. We therefore use the flatness surrogate\cite{wang2025structureinformedlearningflatband} as an enrichment step. Its RMSE of 0.236 on the test set is sufficient for this recall-oriented pre-filter role, allowing likely flat-band candidates to be prioritized while reserving full DFT verification for the most promising structures. Applying a predicted-flatness threshold of 0.5 reduces the pool to 9,352 structures. This subset defines the flat-band-prioritized region of the generated chemical space, comprising structures that are not only chemically plausible after prescreening but also electronically likely to retain the targeted flat-band character.

First-principles calculations confirm that this prioritization strongly enriches the candidate pool in flat-band materials. Of 300 structures randomly sampled from this pool, 256 (85\%) converged, yielding band structures and densities of states, and 184 produced band structures suitable for evaluation with the flatness metric of Ref.~\cite{wang2025structureinformedlearningflatband}. Among these scored structures, 134 exceeded a flatness score of 0.5 and 78 exceeded 0.95. The resulting medium-to-high-flatness rate of 73\% (134/184) is the central quantitative outcome of the pipeline: unconventional skeletons, once expanded into complete crystals, reliably yield flat bands, retrospectively validating the combined generation-and-filtering pipeline. 

The next question is whether these flat bands are caused by the imposed skeletons or arise incidentally from the generated chemistry. We therefore looked for candidates in which two signatures coincide: the DFT flat-band spectral weight is concentrated on the constrained skeleton, and the re-extracted skeleton still supports a corresponding flat band in the nearest-neighbour tight-binding model. When both signatures are present, the flat feature in the full crystal can be traced back to the connectivity of the imposed skeleton, providing a mechanistic link between the geometric constraint and the DFT electronic structure.

\begin{figure}
    \centering
    \includegraphics[width=\linewidth]{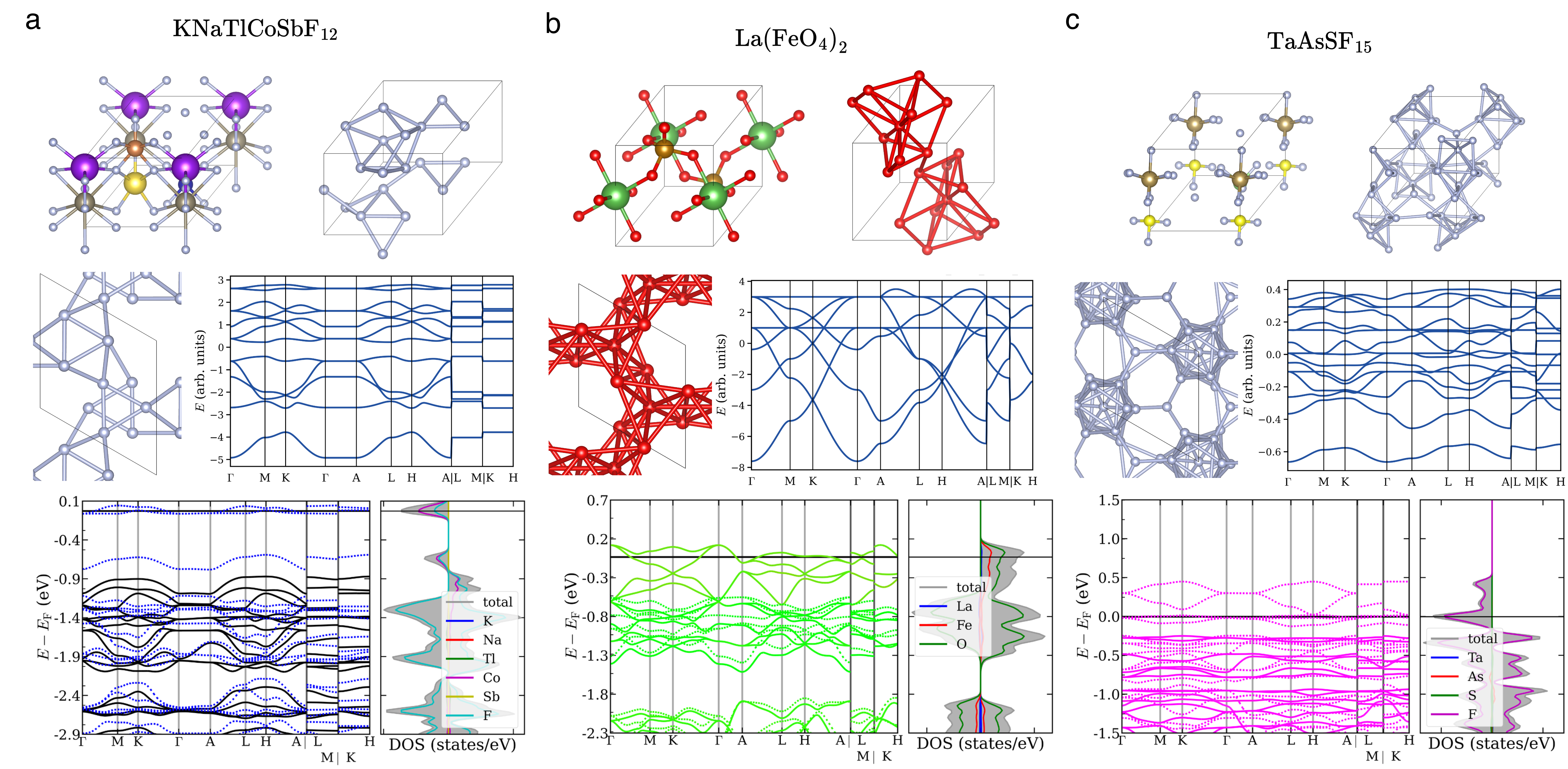}
    \caption{\textbf{Selected showcases of generated flat-band materials.}
    For each candidate, the upper-left panel shows the generated crystal structure; the upper-right panel shows a 3D view of the crystal net that corresponds to the constrained sublattice and produces a flat band in the nearest-neighbour model; the middle-left panel shows a c-direction top view of the same crystal net; the middle-right panel shows the nearest-neighbour tight-binding band structure of the isolated skeleton; and the lower panel shows the DFT band structure with bands projected onto the constituent elements, alongside the corresponding element-projected density of states. All DFT energies are referenced to the Fermi level. In the DFT band structure diagrams, all flat-band regions are placed at the centre of the energy range spanned. \textbf{a}, \ce{KNaTlCoSbF12} with space group 143 (P3). The top view is along the c-axis, perpendicular to the plane of the 2D crystal net. \textbf{b}, \ce{La(FeO4)2} in a trigonal setting with space group 147 (P$\bar{3}$). \textbf{c}, \ce{TaAsSF15} with space group 143 (P3).}
    \label{fig:dft}
\end{figure}

Figure \ref{fig:dft} presents three representative geometry-derived candidates, chosen to illustrate the structural diversity of the unconventional skeletons identified above: a quasi-two-dimensional net extended into a 3D crystal, a corner-sharing tetrahedral framework, and a fully three-dimensional connectivity with no counterpart among named flat-band geometries. In all three cases, band touching points occur at high-symmetry points, showing clear potential for topologically non-trivial flat bands. \ce{KNaTlCoSbF12} (Fig. \ref{fig:dft}\textbf{a}) showcases an F-based skeleton. The crystal net related to the flat band is two-dimensional, similar to the case of the Cs-based sublattice in \ce{Cs2TeO3} mentioned above, representing an extension of a 2D net into a 3D crystal. Its DFT band structure exhibits a flat band around 1.4 eV below the Fermi energy with a shape closely capturing that of the tight-binding result. \ce{La(FeO4)2} (Fig. \ref{fig:dft}\textbf{b}) is built around an O-based skeleton in a corner-sharing tetrahedral crystal net similar to the unconventional sublattice of \ce{SrAl2B2O7} discussed before; the isolated tight-binding model of the O sublattice reproduces a dispersionless band, and the corresponding DFT band structure exhibits a manifold of flat bands 0.8 eV below the Fermi level whose projected DOS is dominated by O with Fe hybridization. \ce{TaAsSF15} (Fig. \ref{fig:dft}\textbf{c}) illustrates the case of a complex three-dimensional F-based skeleton that cannot be referred to any known flat-band structure, similar to the Fe-based sublattice in \ce{Fe7S8}: the tight-binding spectrum of the bare sublattice supports multiple near-zero-energy flat bands, and the full DFT calculation recovers flat features right at the Fermi level. These examples demonstrate that the resulting flat bands are not incidental narrow bands of the surrounding chemistry. Across three distinct classes of unconventional skeletons, the flat features in the DFT calculations remain tied to the tight-binding spectra of the imposed sublattices, while the band touchings at high-symmetry points indicate that the relevant crystallographic symmetry is retained. The result is a mechanism-preserving transfer of flat-band geometry into new chemical environments.

\section*{Discussion}\label{sec13}

Our results show that the geometric origins of flat bands in real compounds are not exhausted by the established catalogue of named motifs: the kagome, Lieb, pyrochlore and related nets chart only a well-explored corner of a much broader landscape of flat-band-generating connectivity. Embedding tight-binding-validated sublattices in this landscape exposes a population of physically grounded skeletons between and beyond the known families, and symmetry-aware generative diffusion renders them as chemically realizable crystals in which the geometric mechanism demonstrably survives the change of chemical environment.


More broadly, our results return flat-band design to its mechanism-first origins -- the tradition of Lieb's bipartite lattices, Mielke's line graphs, Tasaki's decorated lattices, and systematic compact-localized-state generators~\cite{lieb1989two,mielke1991ferromagnetic,tasaki1998nagaoka,maimaiti2017compact, liu2025designing} -- while relocating that logic from abstract lattice models to crystal generation. The difficult step has been realizing such a mechanism in a chemically concrete crystal of specified symmetry, which materials-level efforts have so far approached by screening databases for motifs already known to host flat bands~\cite{regnault2022catalogue,chiu2022line}. Here this model-to-material gap is crossed by generation rather than search. Doing so places a demand on generative models that prototype-based design does not: band touchings and topological character depend on the symmetry of the host crystal, so a skeleton must remain embedded in the crystallographic language of the generated material rather than be preserved as a bare geometric shape. It is precisely this symmetry-resolved embedding that makes uncatalogued flat-band geometries transferable into new chemical environments.

The present implementation deliberately focuses on connectivity-driven flatness captured by nearest-neighbour models, where the link between geometry and electronic structure is most transparent. Extending the skeleton description to include multi-orbital character and spin–orbit coupling would allow the same strategy to address flat bands with orbital symmetry and relativistic effects. Conditioning the generative process on targeted topological invariants, filling fractions, or correlated ground states would move the approach from discovering flat-band candidates to designing the phases they are meant to host. This points to a broader form of mechanism-preserving materials design, in which local electronic mechanisms are not only recognized in existing compounds but transferred into new crystals as symmetry-compatible generative constraints.

\section*{Methods}\label{sec11}
\subsection*{Flat-band sublattice search from Materials Project}
Here we illustrate the workflow of screening for sublattices that contribute to non-trivial flat bands. We adapt the high-throughput method in \cite{wang2025structureinformedlearningflatband} defining a physical score which combines bandwidth and density of states (DOS) to screen 29,465 materials with band structure data available from Materials Project database. 

The bandwidth score $S_{\text{bandwidth}}$ is determined by the bandwidth of the flattest band. The threshold bandwidth $\omega_{\text{max}}$ above which a material is considered not flat-band is 300 meV. The DOS score $S_{\text{DOS}}$ is derived comparing the average DOS in the local window at the flat-band energy and the average DOS within a fixed range $[-5, 5]$ eV. The overall flatness score is computed as
\begin{equation}
\begin{aligned}
S_{\text{total}} = 
\left\{
\begin{array}{ll}
0, & \text{if } \Delta E > \omega_{\text{max}}  \\
\text{sigmoid}\big(\lambda \cdot (S_\text{bandwidth}+S_\text{DOS})\big) \cdot & \\
\end{array}
\right. \\
\quad \text{sigmoid}\big(\beta \cdot (S_\text{bandwidth} \cdot S_\text{DOS})), \quad \text{otherwise}
\end{aligned}
\end{equation}
where $\lambda=5.37$ and $\beta=9.26$, determined via Bayesian Optimization on a physics-motivated target function shaped by HDBSCAN clustering \cite{wang2025structureinformedlearningflatband}. 4,155 materials of flatness scores $ > 0.95$ are shortlisted for further sublattice screening.

The next step is to search for sublattices associated with non-trivial flat bands. Although it is applicable to only look through sublattices corresponding to elements of highest contribution on DOS within the flat-band energy window, to avoid overlook flat bands close to bands of other elements (secondary elements participate as hopping mediators), we study sublattices of all elements in shortlisted flat-band materials. Similar as in \cite{neves2024crystal}, atoms within a distance $\chi d_{\text{NN}}$ are paired as nearest neighbours (NN), where $d_{\text{NN}}$ is the shortest distance between any atom in the sublattice and $\chi$ is chosen to be 1.02, 1.05, 1.1, 1.2 and 1.4. Then we build tight-binding models written as 
\begin{equation*}
    H = -t \sum_{\langle i,j \rangle} c_i^\dagger c_j + c_j^\dagger c_i 
\end{equation*}
where $c_i$($c_i^\dagger$) is fermion annihilation (creation) operator at site $i$ and $t$ is the hopping parameter between nearest neighbour pairs $\langle i,j \rangle$. The band structure is then calculated on $16 \times 16 \times 16$ grid mesh and flat-band is identified by bandwidth estimation. In the above process, isolated groups of atoms (atoms cannot hop infinite far) are eliminated from the tight-binding model to remove trivial flat band.

\subsection*{GNN encoder trained through self-supervised learning}
To distinguish unconventional sublattices from common motifs known to be associated with flat-band, we train a message passing GNN encoder to produce geometric embeddings of sublattices via self-supervised learning. Common crystal GNNs use unit cell structure files as their input. These models typically construct their graphs by applying the minimum image convention, where each atom in the unit cell searches for neighbours within a fixed radial cutoff. Here we use a unique scheme where a $3 \times 3 \times 3$ supercell of the sublattice is used to define a crystal graph $\mathcal{G} = (V, E)$ where $V$ and $E$ denote nodes and edges corresponding to atoms and connections between atoms and their nearest neighbours and next-to-nearest neighbours. The cross-boundary edges wrap around to the opposite side of the supercell, maintaining periodic connectivity. In this way, more degrees of freedom for augmentation are added retaining the geometric features of periodic structures. 

In our model, both node features and the message passing process are $E(3)$ invariant. The node features are relative Cartesian positions defined with respect to a central atom whitin a frame fixed by unit cell, detailed in SI II. and edge attributes are interatomic distances. We write atom Catersian coordinates as $x$, node and edge representations as $h$ and $e$. $\mathcal{N}(i)$ is the set of connected neighbours of the $i$-th node in the graph. The message passing layer in our model updating node representations through graph convolution is defined as
\begin{equation*}
    h_i^{(0)} = x_i
\end{equation*}
\begin{equation*}
    m_{ij}^{(l)} = W_{\text{edge}}^{(l)} \cdot [h^{(l)}_j\parallel e_{ij}]
\end{equation*}
\begin{equation*}
    h_i^{(l+1)} = \text{ReLU}(W_{\text{node}}^{(l)} \cdot h_i^{(l)} + \oplus_{j\in \mathcal{N}(i)} m_{ij})
\end{equation*}
where $W_{\text{edge}}$ and $W_{\text{node}}$ are trainable weight tensors.

Each sublattice graph $(V_i,E_i)$ is augmented into a perturbed pair of graphs $(\mathcal{G}_i,\mathcal{G}_i^+)$ with supercell size of $3 \times 3 \times 3$. Their atomic coordinates $V$ are randomly shifted by a small amount and their nodes and edges are randomly masked. Because of the supercell design, perturbation is introduced only locally, and the model is able to "fill the gaps" and understand the global geometric features such as symmetry and periodicity through learning. A contrastive learning loss InfoNCE is used, written as
\begin{equation*}
    \mathcal{L} = - \log \frac{\exp{(\text{sim}(z_i,z_i^+)/\tau)}}{\sum^{2N}_{j=1}\mathds{1}_{[j\neq i]}\exp{(\text{sim}(z_i,z_j)/\tau)}}
\end{equation*}
\begin{equation*}
    \text{sim}(z_i,z_j) = \frac{z_i \cdot z_j}{\left\| z_i \right\| \left\| z_j \right\|}
\end{equation*}
where $z_i$ are emdeddings in the concatenated batch, $z_i^+$ is its positive pair, $N$ is the batch size, $\tau$ is the temperature parameter and $\mathds{1}$ is an indicator function that masks out self-comparison. The loss function encourages the model to maximize the similarity between embeddings of augmented pairs and minimize the similarity between embeddings of different sublattices.

To evaluate our model, we adapt the produced embeddings of sublattices for property prediction tasks without fine-tuning. These results imply that the model has learned to encode important underlying structural geometry. Further details about latent interpretation of the GNN encoder are in Supplementary Information II.

\subsection*{Scoring of unconventional sublattices}
We use the self-supervised GNN encoder to yield latent space representations that encode geometric similarity of screened sublattices related to non-trivial flat bands. An unperturbed crystal graph is built from a $3 \times 3 \times 3$ supercell for each sublattice. In total, 1,852 sublattices are fed through the model to produce representations.

RCSR names of conventional types of sublattices are found using barycentric placement implemented by Systre \cite{delgado2003identification}, classified as standard crystal net prototypes. Besides those sublattices successfully classified by Systre, points within high-density regions of latent space are well-represented in the Materials Project database. Those points tend to have commonly known geometric features and need to be excluded as well.

UMAP is used first to reduce the dimension of embeddings to 10. Through the HDBSCAN algorithm \cite{mcinnes2017hdbscan}, reduced embeddings are clustered to locate regions of high density of points. The minimum size of clusters is chosen to be 8, and the minimum number of samples in a neighborhood for a point to be considered as a core point is 4. For the algorithm to be conservative (it does not cluster together far away points), we set the cluster selection epsilon to 0.1.

To quantify the novelty of each sublattice, we define a novelty score as the minimum Euclidean distance in the reduced latent space from a given sublattice embedding to any point that is either classified by Systre or assigned to an HDBSCAN cluster. Sublattices with near-zero scores (smaller than 0.01), corresponding to points residing within or immediately adjacent to known clusters, are first excluded. From the remaining distribution, sublattices above the 50th percentile threshold (0.27) are selected as true novelty candidates for constrained generation.

\subsection*{Sublattice and symmetry constrained material generation}

To translate the isolated, unconventional sublattices into chemically realizable 3D crystal structures, we introduce SkeleGen, which combines the structural constraint framework SCIGEN~\cite{okabe2025structural} with the symmetry-aware diffusion model DiffCSP++~\cite{jiao2024space}. The SCIGEN masking strategy locks a subset of atoms to a predefined geometric skeleton, while DiffCSP++ conducts diffusion in the O(3)-invariant logarithmic space of the lattice matrix and restricts fractional coordinates to Wyckoff position subspaces, jointly guaranteeing structural and symmetry consistency without retraining or fine-tuning the base model.

Within our framework, a periodic crystal structure $\mathcal{M}$ is represented by the lattice matrix $L \in \mathbb{R}^{3\times 3}$, fractional coordinates $F \in [0,1)^{3\times N}$, and one-hot atom types $A \in \mathbb{R}^{h\times N}$. The generation pipeline proceeds in several stages.

\noindent\textbf{Sublattice initialization.} The unconventional sublattices extracted from the Materials Project are converted into rigid structural initializations. The sublattice sites are populated by sampling broadly from transition metals, lanthanides, main-group metals, and non-metals. Lattice angles are inherited from the sublattice, while lattice lengths are determined through sampling of bond-lengths from distributions extracted from Materials Project.

\noindent\textbf{Symmetry assignment.} The space group $G_{\text{sub}}$ of the sublattice is determined using \texttt{pymatgen} \cite{ong2013python} with a generous symmetry tolerance $0.1$. Since adding unconstrained atoms can only preserve or lower the symmetry, the full crystal must belong to a subgroup $G_{\text{full}} \leq G_{\text{sub}}$. We enumerate the subgroups of $G_{\text{sub}}$ through \texttt{Pyxtal} \cite{fredericks2021pyxtal}. One compatible subgroup is randomly selected as $G_{\text{full}}$. 

\noindent\textbf{Wyckoff configuration.} For the chosen $G_{\text{full}}$, the constrained sublattice atoms are matched to Wyckoff positions by finding transformation pairs $\{(R_{s_i}, t_{s_i})\}$ such that each sublattice coordinate satisfies $f_{s_i} = R_{s_i} f'_s + t_{s_i}$ for a basic coordinate $f'_s$, within a tight positional tolerance. This translates the structural position constraint into a Wyckoff position constraint, yielding a partial configuration $W_c$. The remaining Wyckoff positions for unconstrained atoms are completed by a dynamic-programming sampler that proposes geometrically consistent Wyckoff letters and multiplicities satisfying $\sum_i a_i m_i = N - N_c$, where $N_c$ is the number of constrained atoms, details provided in SI IV. Importantly, this sampler determines only the geometric skeleton; the element types of unconstrained atoms are generated by the diffusion process itself.

\noindent\textbf{Constrained diffusion.} Generation proceeds via DiffCSP++'s joint diffusion on the lattice coefficient vector $\mathbf{k}$, basic Wyckoff coordinates $F' \in \mathbb{R}^{3\times N'}$, and basic atom types $A' \in \mathbb{R}^{h\times N'}$, modified to incorporate structural constraints through the SCIGEN masking strategy.

For the lattice, the sublattice fully determines $L$ and therefore $\mathbf{k}$ is entirely fixed; no lattice diffusion occurs and the crystal family constraint is automatically satisfied.

For fractional coordinates, diffusion operates on the $N'$ basic Wyckoff coordinates. Noise is projected onto the Wyckoff subspace via $\epsilon'_{F'}[:,s] = R^\dagger_{s_0} \epsilon_{F'}[:,s]$. A binary constraint mask distinguishes the fixed skeleton ($m=1$) from freely variable unconstrained atoms ($m=0$). At each timestep $t$, the basic coordinates of constrained Wyckoff positions are replaced with values from a predefined forward-diffused sublattice trajectory, while unconstrained positions are denoised normally by the model $\phi(\mathcal{M}_t, t)$.

For atom types, constrained Wyckoff positions have their types fixed to the sublattice element at each step via mask replacement. Unconstrained positions are initialized from the prior $\mathcal{N}(0, I)$ and denoised through the standard DDPM backward process, with WyckoffMean ensuring atoms within the same orbit share the same type.

The complete structure at each timestep is assembled as:
\begin{equation}
    \mathcal{M}_t \leftarrow m \odot \mathcal{M}_t^c + (1-m) \odot \mathcal{M}_t^u
\end{equation}
where $\mathcal{M}_t^c$ follows the predefined constrained trajectory and $\mathcal{M}_t^u$ is produced by the denoising model. This recombination forces the base model to navigate the conditional chemical space, predicting compatible element types and coordinates for unconstrained atoms that satisfy charge neutrality and optimize local coordination environments around the rigid skeleton.

\noindent\textbf{Adaptive unmasking.} In the final $T_{\text{stop}}$ steps ($T_{\text{stop}}/T = 0.1$), constrained atoms transition from mask replacement to standard DiffCSP++ Wyckoff-constrained denoising. Their positions are no longer overwritten by the predetermined trajectory but are instead denoised by the model subject to Wyckoff subspace projection. This allows small positional relaxation within the symmetry-permitted subspace, reconciling residual tension between structural and symmetry constraints while preserving the space group, yielding a more stable final structure $\mathcal{M}_0$ that retains the geometry of the embedded flat-band skeleton.

\subsection*{First-principle simulation}
\textbf{Structural relaxation.} 
To assess the structural stability of candidate models we carry out plane-wave DFT calculations using VASP, adopting a computationally efficient setup for this initial relaxation stage. The PBEsol exchange–correlation functional (GGA framework) is used together with PBE\_54 PAW potentials. The plane-wave kinetic-energy cutoff is set to 1.2 times the largest default cutoff among the constituent elements (ENMAX scaling). Brillouin zone sampling uses a $\Gamma$-centred grid with a linear k-point density of approximately 5 per $\text{nm}^{-1}$.  The electronic self-consistency threshold is $10^{-5}$ eV, with at most 60 SCF iterations. Ionic relaxation (conjugate-gradient, IBRION = 2) proceeds until the maximum force on any atom falls below 0.05 eV/Å, or 150 steps are reached. Cell shape, volume, and all internal coordinates are relaxed simultaneously (ISIF = 3). Spin-polarised calculations are performed (ISPIN = 2) with element-wise initial magnetic moments, enabled automatically when magnetic species are present.

\noindent\textbf{Static calculation.}
Electronic structure is then computed in a three-step sequence using atomate2's BandStructureMaker: a self-consistent static calculation, followed by non‑self‑consistent uniform (DOS) and line-mode (band-structure) runs \cite{ganose2025_atomate2}. PAW potentials, PBEsol, spin polarisation, the LDA+U corrections (Dudarev scheme, with literature $U$ values for the 3d transition metals), and the 1.2 $\times$ plane-wave cutoff are inherited from the relaxation. The static step tightens the SCF threshold to $10^{-6}$ eV (max 100 iterations) and a $\Gamma$-centred grid at $K_{\text{spacing}} = 0.188$ Å$^{-1}$. The resulting charge density is then re-used (ICHARG = 11) for the non-SCF runs, both employing Gaussian smearing of 0.05 eV. The uniform DOS run uses a denser $K_{\text{spacing}} = 0.1$ Å$^{-1}$ $\Gamma$-centred mesh, while the band-structure run samples a high-symmetry path generated by pymatgen's HighSymmKpath at a line density of 20 k‑points per Å$^{-1}$.

\section*{Code Availability}
The code for this study is publicly available at: https://github.com/OHOHTI/FlatGen.
\bibliography{sn-bibliography}

\backmatter


%

\end{document}